\documentclass{amsart}[11pt]
\usepackage{amsmath}
\usepackage{amssymb}
\usepackage{amsthm}
\usepackage{mathrsfs}
\usepackage[margin=1in]{geometry}
\usepackage[hang,small,bf]{caption}
\usepackage{tikz}
\usepackage{braket}
\usetikzlibrary{backgrounds,shadows.blur,fit,decorations.pathreplacing,shapes}
\usepackage{color}
\usepackage{enumerate}
\usepackage{graphicx}
\usepackage{amsfonts}
\usepackage{hyperref}
\usepackage{enumerate}
\usepackage{xcolor}
\usepackage{subfigure}
\usepackage{float}
\usepackage{epstopdf} 
\restylefloat{table}
\newtheorem{theorem}{Theorem}[section]

\theoremstyle{definition}

\newtheorem{assumption}[theorem]{Assumption}

\numberwithin{equation}{section}

\newcommand{\init}{\psi_{\mathrm{init}}}
\newcommand{\out}{\psi_{\mathrm{F}}}

\newcommand{\CC}{\mathbb{C}}
\newcommand{\RR}{\mathbb{R}}
\newcommand{\EE}{\mathbb{E}}
\newcommand{\D}{\mathrm{d}}
\newcommand{\cells}{\mathrm{cells}}
\newcommand{\NN}{\mathbb{N}}
\newcommand{\Hf}{\mathfrak{H}}
\newcommand{\Hh}{\mathcal{H}}

\newcommand{\Mm}{\mathcal{M}}
\newcommand{\Oo}{\mathcal{O}}
\newcommand{\Ss}{\mathcal{S}}
\newcommand{\E}{\mathrm{e}}
\newcommand{\I}{\mathrm{i}}
\newcommand{\uu}{\mathrm{u}}
\newcommand{\vv}{\mathrm{v}}

\newcommand{\Omt}{\mathbf{O}_{2}}
\newcommand{\Imt}{\mathbf{I}_{2}}
\newcommand{\Am}{\mathrm{A}}
\newcommand{\Cm}{\mathrm{C}}
\newcommand{\Um}{\mathrm{U}}
\newcommand{\Ym}{\mathbf{Y}}
\newcommand{\Xm}{\mathbf{X}}
\newcommand{\Rm}{\mathbf{R}}
\newcommand{\Hm}{\mathbf{H}}

\newcommand{\ttheta}{\boldsymbol \theta}

\newcommand{\eps}{\varepsilon}
\newcommand{\Err}{\boldsymbol{\mathcal{E}}}
\DeclareMathOperator*{\argmin}{arg\,min}

\begin{document}

\title{A Quantum algorithm for linear PDEs arising in Finance}

\author{Filipe  Fontanela}
\address{Lloyds Banking Group plc, Commercial Banking, 10 Gresham Street, London, EC2V 7AE,~UK}
\email{Filipe.Fontanela@lloydsbanking.com}

\author{Antoine Jacquier}
\address{Department of Mathematics, Imperial College London, and Alan Turing Institute}
\email{a.jacquier@imperial.ac.uk}

\author{Mugad Oumgari}
\address{Lloyds Banking Group plc, Commercial Banking, 10 Gresham Street, London, EC2V 7AE,~UK}
\email{Mugad.Oumgari@lloydsbanking.com}

\thanks{The views and opinions expressed here are the authors' and do not represent the opinions of their employers. 
They are not responsible for any use that may be made of these contents. No part of this
presentation is intended to influence investment decisions or promote any product or service.
The authors would like to thank Alexei Kondratyev for stimulating discussions.}
\date{\today}

\begin{abstract}
We propose a hybrid quantum-classical algorithm, originated from quantum chemistry, to price European and Asian options
in the Black-Scholes model. 
Our approach is based on the equivalence between the pricing partial differential equation and the Schr\"{o}dinger equation
in imaginary time.
We devise a strategy to build a shallow quantum circuit approximation to this equation, only requiring few qubits.
This constitutes a promising candidate for the application of Quantum Computing techniques (with large number of qubits affected by noise)
in Quantitative Finance.
\end{abstract}

\maketitle


\section{Introduction}

Pricing financial derivatives accurately and efficiently is one of the most difficult and exciting challenges in Mathematical Finance. 
While the existence of closed-form solutions are known in very few simple cases, 
the vast majority of models do not admit one, and computing derivatives rely on computational techniques,
either using Monte Carlo simulation methods or via numerical methods for partial differential equations, such as finite differences 
or finite elements.
Unfortunately, these techniques can rapidly become computationally intensive when the model becomes complicated, 
or when the dimension of the problem (in the case of Basket options for example) becomes large.
While huge progress has been made over the past decades, these techniques have inherent limitations which cannot be overcome 
when run on classical computers.

Recently, the emergence of small-scale quantum computers has caught the eyes of 
mathematicians and financial engineers~\cite{Martin2019, Orus2019}
as a a potential goose laying golden eggs.
Indeed, while Quantum Computing has been around since Benioff~\cite{Benioff}, Deutsch~\cite{Deutsch}, Feynman~\cite{Feynman} and Manin~\cite{Manin} suggested
that a quantum computer could perform tasks out of reach for classical computers, 
it really started getting some traction when Shor~\cite{Shor} unearthed a polynomial-time quantum algorithm to factor integers.
Since then, huge efforts have been made to actually build quantum hardware;
only recently though have researchers, in a partnership between Google~AI and the US NASA, 
managed to stabilise a quantum chip over a short period of time, 
effectively starting what has already been called the Quantum Revolution.
Several companies are now proposing online quantum capabilities, using a few qubits.
We are still far from being able to use those in production mode on a large scale, 
but the revolution is fast marching, and tools are urgently needed to embrace it.
Quantitative Finance is traditionally quick to respond to such calls, 
and several attempts have recently been made to use quantum techniques for option pricing, 
in particular~\cite{Rebentrost2018,Stamatopoulos2019}, 
focusing on the so-called Amplitude Estimation algorithm~\cite{Brassard2002}, 
which allows a quadratic speed-up for simulation methods compared to classical Monte Carlo schemes. 

We propose a new route to investigate the use of Quantum techniques in Quantitative Finance, 
and develop a hybrid quantum-classical algorithm to solve the Schr\"odinger equation,
which by simple transformations is equivalent to the PDE satisfied by the option price.
The paper is organised as follows. 
In Section~\ref{Sec:BS}, we introduce the main tools and notations,
clarifying the link between the Black-Scholes PDE~\cite{BS} and its Schr\"odinger counterpart along the imaginary axis, 
while providing a short reminder on Quantum mechanics.
In Section~\ref{Sec:IT}, we borrow an idea developed in quantum chemistry~\cite{McArdle2019}, 
and propose an algorithm to solve the aforementioned Schr\"{o}dinger equation using a quantum computer. 
The methodology develops a hybrid algorithm, where part of the computations can be run on a quantum computer, 
while the remaining tasks are solved on a classical machine. 
We describe in detail in Section~\ref{Sec:Algo} the actual implementation of the algorithm, 
with a particular emphasis on the quantum circuit. 
The numerical results are presented in Section~\ref{Sec:NE}, and we show that the shallow quantum circuit developed here
is able to price European and arithmetic Asian options in the Black-Scholes model with very good accuracy. 

\section{Black-Scholes and the Schr\"{o}dinger equation} \label{Sec:BS}
The Black-Scholes model is at the core of financial modelling and assumes that, under a given risk-neutral measure, 
the underlying stock price process~$(S_t)_{t\geq 0}$ satisfies the stochastic differential equation
\begin{equation}\label{Eq:BlackScholes}
\frac{\D S_t}{S_t} = r \D t + \sigma \D W_t, \qquad\text{for }t\geq 0,
\end{equation} 
where $W$ is a standard Brownian motion on a given filtered probability space, $r\in\RR$ is the instantaneous risk-free rate, 
and $\sigma>0$ is the instantaneous volatility. 
There is a vast literature extending this model in many different directions and using a plethora of numerical techniques, 
ranging from PDEs to Monte Carlo and Fourier methods. 
We are here chiefly interested in developing a quantum-based algorithm for PDEs, 
and will hence ignore numerical methods apart from the PDE approach, considering only European financial derivatives, 
with no early exercise. 

\subsection{Pricing PDEs}
For a European Call option with payoff $f(S_T) = \max(S_T - K, 0)$ at maturity~$T$, 
the Feynman-Kac formulation allows us to write the option price $V(t,s) := \EE\left[\E^{-r(T-t)}f(S(_T)\vert S_t=s\right]$, for any $t\in [0,T]$, obtained by no-arbitrage arguments, as the unique smooth solution to the PDE
\begin{equation}\label{Eq:BS_1}
	\left(\partial_t + \frac{\sigma^2 s^2}{2}\partial_{ss} +  r S \partial_{s} - r\right)V(t,s) = 0,
\end{equation}
for all $s>0$, $t\in [0, T)$, with terminal condition $V(T,s)=f(s)$. The changes of variables $s=\E^{x}$ and $\tau = \sigma^2(T-t)$ reduce the PDE to:

\begin{eqnarray} \label{Eq:BS_1b}
\left(-\sigma^2{\partial_\tau}+\left(r-\frac{1}{2}\sigma^2\right){\partial_x}+\frac{1}{2}\sigma^2 {\partial_{xx}} -r \right)  V\left(\tau,x\right)=0
\end{eqnarray}
Finally, by applying a second round of changes of variables $V(\tau, x)=e^{a x+ b \tau} u(\tau, x)$,
we get to the heat equation:
\begin{equation}
\label{Eq:Heat}
\partial_{\tau} u(\tau,x) = \frac{1}{2}\partial_{xx}u(\tau, x),
\end{equation}
on $(0,\sigma^2 T]\times\RR$, 
with boundary condition $u(0,x)=\E^{-ax}f(\E^x)$, where with $a=\frac{1}{2}-\frac{r}{\sigma^2}$, $ b=-\frac{1}{2}\left(\frac{1}{2}-\frac{r}{\sigma^2}\right)^2-\frac{r}{\sigma^2}$. Another common derivative is the fixed-strike arithmetic Asian Call option, with payoff
\begin{equation}\label{Eq:POA}
\max\left( \frac{1}{T}\int_0^T S_{t} \D t - K, 0\right),
\end{equation} 
for some strike $K>0$. 
Setting 
$Y_{t} := \left(B_{t} - K\E^{-r(T - t)}\right) / S_{t}$
with 
\begin{equation*}
B_{t} = q(t) S_{t} + \frac{\E^{-r(T-t)}}{T}\int_0^t S_{u}\D u
\qquad\text{and}\qquad
q(t) = 
\left\{
\begin{array}{ll}
\displaystyle \frac{1}{rT}\left(1-\E^{-r(T-t)}\right),  & \mbox{if } r\ne 0,\\
\displaystyle 1-\frac{t}{T},  & \mbox{if } r=0,
\end{array}
\right.
\end{equation*}
Brown~\cite{Brown2016} and Vecer~\cite{Vecer2001} showed that the price at time $t\in [0,T)$ of the option reads
$V(t, S_{t}) = S_{t} \widetilde{Q}(t,Y_{t})$;
with $\tau:=\sigma^2(T-t)$ and $Q(\tau, y) = \widetilde{Q}(t,y)$, the function~$Q$ uniquely solves  the PDE
\begin{equation}\label{Eq:Vecer}
\partial_{\tau} Q(\tau, y) =  \frac{\left(q(\tau) - y\right)^2}{2}\partial_{yy} Q(\tau, y),
\end{equation}
for $(\tau,y) \in (0,\sigma^2 T]\times\RR$ with terminal condition $Q(0,y) = \max(y,0)$, 
while the space boundaries are set as $\lim_{y\downarrow-\infty}Q(\tau,y)=0$ and $Q(\tau, q(0)) = q(0)$. 
One should note that~\eqref{Eq:Vecer} is similar to the heat equation~\eqref{Eq:Heat}, 
albeit with an additional time-dependence on the right-hand side.

\subsection{Schr\"{o}dinger's formulation}\label{sec:Schrodinger}
Similarly to~\cite{Contreras}, we now translate the pricing PDEs~\eqref{Eq:Heat} and~\eqref{Eq:Vecer}
into linear Schr\"{o}dinger equations, classical  in quantum mechanics. 
The Wick rotation $\xi = -\I\tau$ transforms~\eqref{Eq:Heat} into
$-\I \partial_{\xi}u(\xi, x) = \frac{1}{2} \partial_{xx}u(\xi,x)$,
or, in the quantum computing language using Dirac's notation (Section~\ref{sec:QuantumNotes}), 
\begin{equation}\label{Eq:SE_2}
-\I\frac{\partial}{\partial \xi}\ket{\psi} = \widehat{\Hh} \ket{\psi},
\end{equation}
where the wave function~$\psi$ (and its associated quantum state~$\ket{\psi}$) 
plays the role of the modified price~$u(\cdot, \cdot)$, 
while the Hamiltonian operator is here $\widehat{\Hh}=\frac{1}{2}\partial_{xx}$. 
For this special case of a time-independent Hamiltonian, the solution to~\eqref{Eq:SE_2} is  given explicitly by
\begin{equation}\label{Eq:DynEvol}
\ket{\psi(\xi)} = \exp\left(\I\widehat{\Hh}\xi\right) \ket{\psi(0)},
\end{equation} 
where $\exp\left(\I\widehat{\Hh}\xi\right)$ is the time evolution operator
and $\ket{\psi(0)}$ a normalised initial state with $\braket{\psi(0)|\psi(0)}=1$.
For Asian options, the Wick rotation $\xi = -\I\tau$ now turns~\eqref{Eq:Vecer} into
$$
-\I \partial_{\xi}Q(\tau, y) =  \frac{\left(q(\tau) - Y \right)^2}{2} \partial_{yy}Q(\tau, y),
$$
or, in Dirac's notations, 
\begin{equation}\label{Eq:SE_4}
-\I\frac{\partial}{\partial \xi}\ket{\psi} = \widehat{\Hh}(\xi) \ket{\psi}.
\end{equation}
The main difference between ~\eqref{Eq:SE_2} and~\eqref{Eq:SE_4} is the time-dependent Hamiltonian.	
Nonetheless, the solution can be written by computing the new evolution operator as 
\begin{equation}\label{Eq:DynEvol2}
\ket{\psi(\xi)} = \exp\left(\I\int_0^\xi \widehat{\Hh}(\chi)\D\chi\right)\ket{\psi(0)}.
\end{equation}

\subsection{Reminder on quantum notations}\label{sec:QuantumNotes}
To facilitate the integration of Quantum Mechanics into the realm of Quantitative Finance, 
we shall endeavour to combine notations from both fields in a clear and consistent manner.
Following Dirac's approach, in a given (complex-valued) Hilbert space~$\Hf$, a vector~$\vv$ is represented via the ket notation~$\ket{\vv}\in\Hf$.
For $\uu\in\Hf$, the bra~$\bra{\uu}$ belongs to the dual space~$\Hf^*$, so is a linear map from~$\Hf$ to~$\CC$, 
and we denote the action of~$\uu$ on $\vv$ as the bracket $\braket{\uu|\vv}$.
In classical linear algebra notations, 
suppose that~$\Hf$ is of dimension~$n\in\NN$, then $\uu, \vv\in\Hf$ can be represented as vectors in~$\RR^n$ as
$$
\uu = \begin{pmatrix}
u_1 \\ \vdots \\ u_n
\end{pmatrix}
\qquad\text{and}\qquad
\vv = \begin{pmatrix}
v_1 \\ \vdots \\ v_n
\end{pmatrix},
$$
or $\uu = (u_1, \ldots, u_n)^\top$ and $\vv = (v_1, \ldots, v_n)^\top$.
In that case, we can write
$$
\ket{\vv} = \begin{pmatrix}
v_1 \\ \vdots \\ v_n
\end{pmatrix},
\qquad
\bra{\uu} = \left(u_1^*, \ldots, u_n^*\right),
\qquad\text{and}\qquad
\braket{\uu|\vv} = \sum_{i=1}^{n}u_i^* v_i,
$$
where~$^*$ denotes complex conjugacy.
A physical state in quantum mechanics is represented by a state vector, or ket.
The general quantum state of a qubit, the basic unit of quantum information, is a linear superposition of its orthonormal basis.
In particular, a single qubit state reads $\ket{\psi} = \alpha \ket{0} + \beta\ket{1}$, for $\alpha, \beta\in\CC$ 
satisfying $|\alpha|^2+|\beta|^2=1$, 
with $(\ket{0}, \ket{1})$ the orthonormal basis. 
In classical linear algebra notations, this usually reads
$$
\psi = \begin{pmatrix}
\alpha \\ \beta
\end{pmatrix}
= \alpha\begin{pmatrix}
1\\0
\end{pmatrix}
+\beta\begin{pmatrix}
0\\1
\end{pmatrix}.
$$
Generally speaking, an n-qubit quantum state corresponds to a vector in $\CC^{2^n}$.
For such a state~$\ket{\psi}$, that is, a physical system which can be in~$n$ different, mutually exclusive classical states
$\ket{0}, \ket{1}, \ldots, \ket{n-1}$, so that
$$
\ket{\psi} = \alpha_0 \ket{0} + \cdots + \alpha_{n-1}\ket{n-1},
$$
for $(\alpha_0, \ldots, \alpha_{n-1})\in\CC^{n}$, 
such that $\sum_{i=1}^{n}|\alpha_i|^2 = 1$.
The basis $(\ket{0}, \ket{1}, \ldots, \ket{n-1})$ forms an orthonormal basis of an $n$-dimensional Hilbert space~$\Hf_n$.
Given~$\Hf_n$ and~$\Hf_m$, we can define the tensor product $\Hf:=\Hf_n \otimes \Hf_m$ as the $nm$-dimensional Hilbert space 
spanned by $\{\ket{i}\otimes\ket{j}: i=0,\ldots, n-1, j=0,\ldots, m-1\}$.
For example, a 2-qubit system, corresponding to a Hilbert space of dimension~$4$ can be viewed as 
the tensor product of two Hilbert spaces, each of dimension~$2$, and its basis being spanned by 
$\{\ket{0}\otimes\ket{0}, \ket{0}\otimes\ket{1}, \ket{1}\otimes\ket{0}, \ket{1}\otimes\ket{1}\}$, 
usually written in the denser form $\{\ket{00}, \ket{01}, \ket{10}, \ket{11}\}$.

Quantum logic gates are reversible quantum circuits operating on quantum states, and are represented as unitary matrices.
We shall here make use of several particular gates repeatedly, which can be easily represented in a $1$-qubit state as 
the following matrices in~$\Mm_2$ (the set of  square $2\times 2$ matrices in~$\CC$):
\begin{equation}\label{eq:Gates}
\begin{array}{llc}
\Xm: & \text{X-Pauli gate: } & 
\begin{pmatrix}
0 & 1\\1 & 0
\end{pmatrix}
\\
\Ym: & \text{Y-Pauli gate: } & 
\begin{pmatrix}
0 & -\I \\ \I & 0
\end{pmatrix}
\\
\Hm: & \text{Hadamard gate: } & 
\displaystyle \frac{1}{\sqrt{2}}
\begin{pmatrix}
1 & 1\\1 & -1
\end{pmatrix}
\\
\Rm_{y}(\theta): & \text{Rotation gate with angle~$\theta$: } & 
\displaystyle \begin{pmatrix}
\cos\left(\frac{\theta}{2}\right) & -\sin\left(\frac{\theta}{2}\right)\\
\sin\left(\frac{\theta}{2}\right) & \cos\left(\frac{\theta}{2}\right)
\end{pmatrix}
\end{array}
\end{equation}

\section{Quantum Imaginary Time Evolution} \label{Sec:IT}

We now describe a strategy to solve the linear Schr\"{o}dinger equations~\eqref{Eq:SE_2} and~\eqref{Eq:SE_4} 
along the imaginary time axis using a quantum computer. 
Our approach relies on the hybrid algorithm developed in~\cite{McArdle2019}, 
where part of the computation is performed on a quantum computer, and part on a classical machine. 
Consider first a time-independent Hamiltonian~$\widehat{\Hh}$ with evolution operator (or propagator)~$\exp(\I\widehat{\Hh}\xi)$ 
evolving along real values~$\xi$, as in~\eqref{Eq:DynEvol}. 
In this case, the propagator can be implemented efficiently on a quantum computer by means of a Trotter decomposition, 
since $\exp(\I\widehat{\Hh}\xi)$ is a unitary matrix~\cite{Ying2017}. 
Along the imaginary axis however, the corresponding evolution operator~$\exp(\widehat{\Hh}\tau)$ 
is represented by a non-unitary matrix;
It can be easily simulated on a classical computer but this becomes rapidly infeasible as the dimension of the wave function grows exponentially;
its Trotter decomposition using unitary gates is not straightforward, making its implementation on a quantum computer more challenging.
We follow instead a recent idea~\cite{McArdle2019} to solve an equivalent normalised imaginary time evolution  
\begin{equation}\label{Eq:Im_Time}
\ket{\psi(\tau)} = \gamma(\tau)\  \E^{-\widehat{\Hh}\tau}\ket{\psi(0)},
\qquad\text{with }
\gamma(\tau) := \left(\bra{\psi(0)}\E^{2\widehat{\Hh}\tau}\ket{\psi(0)}\right)^{-1/2},
\end{equation} 
indirectly. 
The parameter~$\gamma(\tau)$ is a normalisation constant and~$\widehat{\Hh}$ a time-independent Hamiltonian. 
We call this approach indirect since~$\ket{\psi(\tau)}$ is not computed by constructing the operator~$\exp(\widehat{\Hh}\tau)$ 
on a quantum computer. 
Instead, we approximate~$\ket{\psi(\tau)}$ by a quantum circuit composed of a sequence of parameterised gates (or trial state)
such that $\ket{\psi(\tau)}\approx\ket{\phi(\ttheta_{\tau})}$, 
where $\ttheta_{\tau} = (\theta_{\tau}^{1},\cdots,\theta_{\tau}^{N})^\top\in\RR^N$ is a vector of time-dependent parameters. 
Knowing how to describe the evolution of~$\ttheta_{\tau}$ thus makes it possible to reconstruct the imaginary Schr\"{o}dinger time evolution $\ket{\psi(\tau)}$ in~\eqref{Eq:Im_Time}. 
We refer to the approximation $\ket{\phi(\ttheta_{\tau})}$ as the ansatz circuit. 
Assume that the initial quantum state is $\ket{\init}$, 
so that the ansatz is $\ket{\phi(\ttheta_{\tau_0})} = \Phi(\ttheta_{\tau_0})\ket{\init}$ at time~$\tau_0$, 
where~$\Phi(\ttheta_{\tau_0})$ is sequence of unitary gates 
$\Phi(\ttheta_{\tau_0}) = \Ss\left(\Um_N(\theta^N_{\tau_0}), \ldots, \Um_k(\theta^k_{\tau_0}), \ldots, \Um_1(\theta^1_{\tau_0})\right)$
which we will specify later. 
We only consider here rotations or controlled rotations for each parameterised gate $\Um_k(\theta^k)$. 
To determine the optimal time-dependent parameters~$\ttheta_{\tau}$--and thus the  Schr\"{o}dinger dynamics--we
minimise the distance $\|\ket{\psi(\tau)} - \ket{\phi(\ttheta_{\tau})}\|$, 
or equivalently solve McLachlan's variational principle~\cite{McLachlan1964}
\begin{equation}
	\delta \left\| \left(\partial_{\tau} + \widehat{\Hh} \right) \ket{\psi(\tau)} \right\| = 0,
	\label{Eq:MLPrinciple}
\end{equation}
where $\left\| v \right\| := \braket{v|v}$, and~$\delta$ denotes infinitesimal variation.
One could alternatively consider the Dirac-Frenkel variational principle~\cite{Dirac, Frenkel}, but, as shown in~\cite{Benjamin-Theory}, 
these variational principles all result in the system of ODEs
\begin{equation}\label{Eq:LinSys}
\Am(\tau)\dot{\ttheta}_\tau = \Cm(\tau),
\end{equation}
for all $\tau$, where $\dot{\ttheta}_\tau:=\partial_\tau\ttheta_\tau$, and the matrix~$\Am(\tau)\in\Mm_N$ and the vector~$\Cm(\tau)\in\RR^N$ read
$$
\Am (\tau) = \left( \Re{\left( \frac{\partial \bra{\phi(\tau)}}{\partial \theta^i}\frac{\partial \ket{\phi(\tau)}}{\partial \theta^j} \right)}\right)_{i,j=1, \ldots,N}
\qquad\text{and}\qquad
\Cm(\tau) = \left(\Re{ \left(  \frac{\partial \bra{\phi(\tau)}}{\partial \theta^i}\widehat{\Hh}\ket{\phi(\tau)}\right)}\right)_{i=1, \ldots,N}.
$$
In this setting, both~$\Am$ and~$\Cm$ can be measured efficiently using a quantum computer~\cite{Ying2017, McArdle2019}. 
In order to build the hybrid classical-quantum scheme, we rely on the following assumptions:

\begin{assumption}\label{Assumptions}\ 
\begin{enumerate}[(i)]
\item Every unitary gate in the algorithm depends on a single parameter.
\item The decomposition
$
\widehat{\Hh}=\sum_{i=1}^{N}\lambda_i h_i
$
holds, for real numbers~$\lambda_i$ and tensor products~$h_i$ of Pauli matrices.
\end{enumerate}
\end{assumption}
Assumption~\ref{Assumptions}(i) is purely for convenience, 
as unitary gates with multiple parameters can be decomposed into single-parameter ones, 
but it allows us to write the derivative of the unitary gates
$$
\partial_{\theta^k} \Um_k(\theta^k) =\sum_{i=1}^{N} f_{k,i}\Um_k(\theta^k) \sigma_{k,i},
\qquad\text{for every } k=1\ldots, N,
$$
where $\sigma_{k,i}$ are one-qubit or two-qubit unitary operators and $f_{k,i}$ a scalar parameter, so that
$$
\partial_{\theta^k} \ket{\phi(\tau)} = \sum_{i=1}^{N} f_{k,i} \widetilde{\Phi}_{k,i} \ket{\init}, 
\qquad\text{where}\qquad
\widetilde{\Phi}_{k,i} = \Ss\left(\Um_N(\theta^N), \ldots, \Um_k(\theta^k)\sigma_{k,i}, \ldots, \Um_1(\theta^1)\right).
$$
For example, if $\Um_k(\theta^k)$ is a single-qubit $\Rm_y(\theta^k)$ rotation gate~\eqref{eq:Gates} such that 
$\Um_k(\theta^k)=\exp\left(-\frac{1}{2}\I\theta^k \sigma_Y\right)$, 
where~$\sigma_Y$ is the Y-Pauli matrix, its derivative is simply 
$\partial_{\theta^k} \Um_k(\theta^k) = -\frac{\I\sigma_Y}{2}\Um_k(\theta^k)$. 
Therefore, the state $\partial_{\theta^k}\ket{\phi(\tau)}$ can be prepared by adding the extra~$\sigma_Y$ gate, 
together with a constant factor $-\frac{\I}{2}$ only, 
such that $\partial_{\theta^k}\ket{\phi(\tau)} = -\frac{\I}{2}\widetilde{\Phi}_{k,k}(\ttheta)\ket{\init}$, 
where $\widetilde{\Phi}_{k,k}(\ttheta) = \Ss(\Um_N(\theta^N),\ldots,\Um_k(\theta^k)\sigma_Y,\cdots,\Um_1(\theta^1))$. 
In general, the matrix~$\Am(\tau)$ is then computed as~\cite{McArdle2019}
\begin{equation}
\Am_{ij}(\tau) = \Re{\left( \sum_{k,l=1, \ldots,N} f_{k,i}^*f_{l,j}\bra{\init}\widetilde{\Phi}_{k,i}^\dagger\widetilde{\Phi}_{l,j}\ket{\init} \right)},
\qquad\text{for }i,j=1, \ldots,N,
\end{equation}
where the dagger superscript~$^\dagger$ denotes the complex conjugate of the transpose, 
and a quantum circuit designed for~$\Am(\tau)$ is presented in~\cite[Appendix]{McArdle2019}
as well as in Section~\ref{Sec:Algo} below.
For the vector~$\Cm(\tau)$, we use Assumption~\ref{Assumptions}(ii); 
this decomposition usually scales polynomially with the system size
since~$\widehat{\Hh}$ is a sparse matrix obtained from the discretisation of a differential operator. 
The corresponding vector~$\Cm$ then reads~\cite{McArdle2019}
$$
\Cm_i(\tau) = \Re{\left( \sum_{k,j=1, \ldots,N}f_{k,i}^*\lambda_j \bra{\init}\widetilde{\Phi}_{k,i}^{\dagger} h_j \Phi \ket{\init} \right)},
\qquad\text{for } i=1, \ldots,N.
$$
Once~$\Am(\tau)$ and~$\Cm(\tau)$ are obtained, the time evolution can be computed numerically using a classical computer. 
We suggest an Euler scheme, so that the evolution of~$\ttheta_{\tau+\Delta_{\tau}}$ is calculated using~\eqref{Eq:LinSys} as
\begin{equation}\label{Eq:Eu_evo}
\ttheta_{\tau+\Delta_{\tau}} = \ttheta_{\tau} + \Delta_{\tau} \dot{\ttheta}_{\tau} = \ttheta_{\tau} + \Delta_{\tau}\Am(\tau)^{-1}\Cm(\tau),
\end{equation}
for some small time step~$\Delta_\tau$.
Usually, the matrix $\Am(\tau)$ is not well-conditioned, and we solve, for each~$\tau$,
\begin{equation}\label{Eq:ThetaDot}
\argmin\left\{ \left\| \Am(\tau)\dot{\ttheta}_\tau - \Cm(\tau) \right\|: \ttheta_\tau\in\RR^N\right\}, 
\end{equation} 
instead of~\eqref{Eq:LinSys} directly, assuming some small cut-off ratio for the eigenvalues of~$\Am(\tau)$;
this is straightforward with the \texttt{Python} package \texttt{numpy.linalg.lstsq} for example. 
For a time-dependent Hamiltonian, for Asian options, we are interested in determining the quantum state
\begin{equation}\label{Eq:HamTime}
\ket{\psi(\tau)} = \gamma(\tau)\  \exp\left(\int_0^\tau \widehat{\Hh}(\chi) \D\chi\right)\ket{\psi(0)},
\quad\text{with }
\gamma(\tau) := \left(\bra{\psi(0)}\exp\left\{2\int_0^\tau \widehat{\Hh}(\chi)\D\chi\right\}\ket{\psi(0)}\right)^{-\frac{1}{2}}
\end{equation}
instead of~\eqref{Eq:Im_Time}, with the new normalisation constant~$\gamma(\tau)$. 
In order to avoid the integral calculation in~\eqref{Eq:HamTime}, 
we use, along  a left-point discretisation of the integral, freezing the time-dependent Hamiltonian~$\widehat{\Hh}(\tau)$ on every time interval;
thus the only difference with the time-independent framework above is the update of the Hamiltonian at every time step.
The main drawback of the present approach is the requirement for an ansatz circuit. 
Ideally, the latter has to be complex enough to approximate the underlying quantum state~$\ket{\psi(\tau)}$, 
but not unnecessarily deep in order to avoid the expensive computation (on a classical computer) of the linear problem~\eqref{Eq:ThetaDot}, i.e. too many parameters~$\ttheta$.
In the examples below, we design a quantum circuit strategy for simple financial derivatives, 
and obtain promising results for relatively shallow quantum circuits. 

\section{Hybrid Quantum-Classical Algorithm} \label{Sec:Algo}
\begin{figure}[h!]
\begin{center}
	\includegraphics[scale=1]{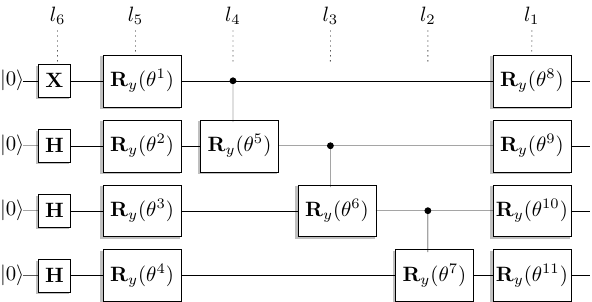}
\end{center}
\caption{Ansatz implemented for European and Asian Call options. 
The vector~$\ttheta$ is composed by the angles of each rotation gate. 
The notations~$l_1, \ldots, l_6$ indicate the different layers.} 
	\label{Fig:AnsatzCircuit}
\end{figure}

We now design the quantum imaginary time algorithm described in Section~\ref{Sec:IT} to price financial derivatives. 
The main challenge is to design an ansatz circuit able to represent the value of the option from expiry down to valuation date. 
We employ the sequence of gates described by the quantum circuit depicted in Figure~\ref{Fig:AnsatzCircuit}.
There, a line represents a qubit, whereas a squared box denotes the action of a quantum gate on the qubit 
on this line. The circuit is to be read from left (inputs) to right (outputs).
A controlled gate (for example~$\Rm_{y}(\theta^5)$) is shown with a filled black circle on the control qubit line, 
and  the usual symbol for the gate is written on the target qubit, with a line connecting the two.
The quantum circuit has 4 qubits only, and thus our strategy is able to recover Call prices with a resolution of 2$^4=16$ points. 
The ansatz in Figure~\ref{Fig:AnsatzCircuit} has a layer~$l_6$ composed by~$\Hm$ and $\Xm$ gates~\eqref{eq:Gates}. 
In the next layer~$l_5$, a sequence of~$\Rm_y$ gates is applied to each qubit. 
Each such gate has a free parameter that needs to be calculated given some desired initial condition, 
and these parameters vary in time, as described above. 
The layers~$l_4$, $l_3$ and~$l_2$, composed by controlled $\Rm_y^c$ gates, and a new layer~$l_1$, similar to~$l_5$, 
are added in order to give more flexibility to the approximation, 
and thus achieve different and more involved values of $\ket{\phi(\ttheta_{\tau})}$. 
In the present analysis, we increase or reduce the complexity of the ansatz circuit by adding or removing new unit cells, 
i.e. layers~$l_4$, $l_3$, $l_2$ and~$l_1$. 
The optimal configuration is obtained when $\ket{\phi(\ttheta_{0}}$ is able to represent $\ket{\psi(0)}$ with good accuracy, 
and with as few free parameters as possible. 
One should note that the ansatz is composed of gates with rotations along the~$y$ axis only. 
The motivation for using~$\Rm_y$ gates is due to its real-valued representation, 
which is desirable as $\ket{\psi(\tau)}$ in~\eqref{Eq:Im_Time} is itself real. 
However, other approaches might also be valid, using for instance~$\Rm_x$ and~$\Rm_z$ gates, 
at the cost of computing the real part of $\ket{\phi(\ttheta_{\tau})}$ at the end of the simulations.

Mathematically, a quantum circuit is described  by means of a series of matrix operations;
the ansatz in Figure~\ref{Fig:AnsatzCircuit} is composed of $\Xm$, $\Hm$, and $\Rm_y$ single-qubits gates, and their mathematical representations are given by the matrices in~\eqref{eq:Gates}.
The controlled rotation gate~$\Rm_y^c$ is implemented using two qubits, and its matrix representation reads
$$
\Rm_y^c(\theta) := 
\begin{pmatrix}
1 & 0 & 0 & 0 \\ 
0 & 1 & 0 & 0 \\ 
0 & 0 &  \cos\left( \frac{\theta}{2}\right)  & -\sin\left( \frac{\theta}{2}\right)  \\ 
0 & 0 & \sin\left( \frac{\theta}{2}\right) &  \ \ \ \cos\left( \frac{\theta}{2}\right)
\end{pmatrix}
 = 
\begin{pmatrix}
\Imt & \Omt \\ 
\Omt & \Rm_y(\theta)
\end{pmatrix},
$$
where~$\Omt$ and~$\Imt$ denote the null and identity matrices in~$\Mm_2$.
The resulting mathematical representation of the quantum circuit in Figure~\ref{Fig:AnsatzCircuit} is calculated by grouping all quantum gates accordingly. 
For example, the effect of the layer $l_1$, which contains the gates $\Rm_y(\theta^{8})$, $\Rm_y(\theta^{9})$, $\Rm_y(\theta^{10})$ and $\Rm_y(\theta^{11})$, is computed as 
$$
\Err^{l_1} := \Rm_y(\theta^{8}) \otimes \Rm_y(\theta^{9}) \otimes \Rm_y(\theta^{10}) \otimes \Rm_y(\theta^{11}),
$$
where $\otimes$ denotes the Kronecker product. 
The effect of the controlled gate~$\Rm_y^c(\theta^{7})$ in $l_2$ is calculated as  
$$
\Err^{l_2} := \Imt \otimes \Imt \otimes \Rm_y^c(\theta^{7}),
$$
and the combined effect of $\Err^{l_1}$ and $\Err^{l_2}$ is the matrix multiplication 
$\Err^{l_1, l_2} :=  \Err^{l_1} \times  \Err^{l_2}$.
This process is repeated through the whole quantum circuit, 
and the resulting effect of the six layers in Figure~\ref{Fig:AnsatzCircuit} reads
$$
\Err^{l_1,\ldots,l_6} = \prod_{i=1}^{6} \Err^{l_i}.
$$
The final output $\ket{\out}$ of the quantum circuit due to the input~$\ket{\init}$ is then computed as 
$$
\ket{\out} = \Err^{l_1, \ldots, l_6} \ket{\init}.
$$
For example, the input $\ket{\init}$ of the ansatz circuit in Figure~\ref{Fig:AnsatzCircuit} is a quantum state 
where all four qubits are~$\ket{0}$, so that its vectorial representation is 
$\ket{\init} = \ket{0} \otimes \ket{0} \otimes \ket{0} \otimes \ket{0} = \ket{0000}$.
Finally, note that if $\ttheta=0$, the resulting state $\ket{\phi(\ttheta)}$ becomes a step function, 
since each gate $\Rm_y$ behaves as the identity operator~$\Imt$.
As discussed above, the payoff of each financial product needs to be represented by the ansatz~$\ket{\phi(\ttheta_{0})}$ before starting the actual simulation. 
In practice, this representation is implemented by computing the values of~$\ttheta_{0}$ which best describes the initial conditions of the algorithm, that is 
\begin{equation}\label{Eq:OptPr}
\ttheta_{0} = \argmin_{\ttheta\in\RR^N} \left\lbrace \left\| \ket{\phi(\ttheta)} - \ket{\psi(0)}\right\| \right\rbrace .
\end{equation}
The quantum state~$\ket{\psi(0)}$ is calculated directly from the payoff~$f(\cdot)$ of each financial derivative. 
In the case of a European option, this is simply
$\ket{\psi(0)}\equiv \gamma(0)\E^{-aX}f(\E^X)$, where~$\gamma(0)$ is the normalisation constant guaranteeing $\braket{\psi(0)|\psi(0)}=1$,
and~$a$ arises in the change of variables from the Black-Scholes PDE to the heat equation~\eqref{Eq:Heat}. 
Some care should be taken when solving~\eqref{Eq:OptPr} since the cost function is not convex, 
and algorithms such as Differential Evolution will avoid being trapped in a local minimum.
Ideally, the design of the ansatz should be considered together with the optimisation problem~\eqref{Eq:OptPr}. 
In practice, if the ansatz circuit is not able to represent the payoff $\ket{\psi(0)}$ accurately, it becomes pointless to compute the time-marching simulation $\ket{\phi(\ttheta_{\tau})}$ since the algorithm already starts from the wrong initial conditions. In this case, the quantum circuit in Figure~\ref{Fig:AnsatzCircuit} needs to be improved for the desired application. The complete strategy to define the depth of the ansatz in Figure~\ref{Fig:AnsatzCircuit} and its initial condition $\ttheta_{0}$ is summarised in Table~\ref{Tab:Algo1}.

\begin{table}[h!]
	\centering
	\begin{tabular}{ll}
		\hline
		\hline
		\textbf{Ansatz and initial condition algorithm}\\
		\hline
		\hline
		\textbf{input} $\leftarrow$ Payoff function & \\
		\qquad{\it initialisation:} \\ 
		\qquad\quad\vline \quad Define the normalised initial condition $\ket{\psi(0)}$ \\
		\qquad\quad\vline \quad Define the maximum acceptable number of unit cells $N_{\cells}^{\max}$ \\
		\qquad\quad\vline \quad Define the maximum acceptable error $\eps^{\max}$=$\|\ket{\phi(\ttheta_{0})} - \ket{\psi(0)}\|$ \\
		\qquad\quad\vline \quad Define an initial ansatz with $N_{\cells}$ unit cells\\	
		\qquad{\it end} \\
		\qquad{\it while $N_{\cells}<N_{\cells}^{\max}$} \\	
		\qquad\quad\vline \quad Compute the optimal $\ttheta_{0}$ from~\eqref{Eq:OptPr} 
and the approximation error $\eps = \| \ket{\phi(\ttheta_{0})} - \ket{\psi(0)}\|$ \\
		\qquad\quad\vline \quad {\it if} $\eps>\eps^{\max}$\\	
		\qquad\quad\vline \qquad\vline \qquad  Increase the number of unit cells $N_{\cells}$ in the ansatz \\
		\qquad\quad\vline \quad {\it else}\\
		\qquad\quad\vline \qquad\vline \qquad  \textbf{output} $\rightarrow$ Ansatz circuit and initial condition $\ttheta_{0}$ \\		
		\qquad\quad\vline \quad {\it end} \\
		\qquad {\it end}\\
		\qquad \textbf{output} $\rightarrow$ Inform that the algorithm has not converged and that a 
different ansatz is needed \\	
		\hline
	\end{tabular}
	\caption{Algorithm to determine the depth of the ansatz and the initial conditions $\ttheta_{0}$.}
	\label{Tab:Algo1}
\end{table}

\begin{figure}[h!]
\begin{center}
	\includegraphics[scale=1]{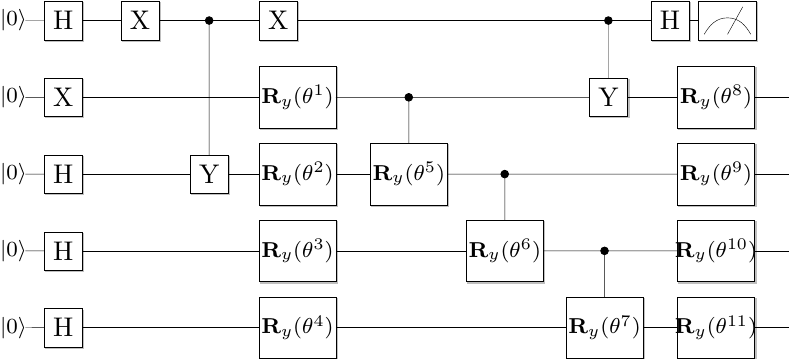}
\end{center}
	\caption{Quantum circuit able to measure $\Am_{2,8}$ in~\eqref{Eq:LinSys}.
The symbol on the top right of the Figure denotes measurement of the quantum state.} 
	\label{Fig:AnsatzMeasurement}
\end{figure}

The main advantage of the hybrid scheme is that both~$\Am(\tau)$ and~$\Cm(\tau)$ in~\eqref{Eq:LinSys} 
can be measured on a quantum computer. 
For example, $\Am_{2,8}(\tau)$ can be measured using the quantum circuit in Figure~\ref{Fig:AnsatzMeasurement}. 

In summary, we need to add an ancillary qubit to the quantum circuit of Figure~\ref{Fig:AnsatzCircuit} together with two gates~$\Hm$ and~$\Xm$,
 and a controlled $\Rm_y^c$ gate before each respective gate $\Rm_y(\theta^2)$ and $\Rm_y(\theta^{8})$. 
In the following, $\Am_{2,8}(\tau)$ is obtained by measuring the expectation value 
on the added ancillary qubit, as in~\cite{McArdle2019}. 
The value of~$\Cm(\tau)$ is similarly implemented on a quantum computer. 
The main difference is that the underlying Hamiltonian~$\widehat{\Hh}$ first needs to be decomposed 
according to Assumption~\ref{Assumptions}(ii).
Table~\ref{Tab:Algo2} below summarises the structure of the hybrid quantum-classical algorithm, 
where the labels (CC) and (QC) indicate which part of the algorithm should be computed on a classical and on a quantum computer respectively.
\begin{table}[h!]
	\centering
	\begin{tabular}{ll}
		\hline
		\hline
		\textbf{Hybrid quantum-classical algorithm} & \\
		\hline
		\hline
		\textbf{input} $\leftarrow$ Ansatz circuit and initial condition $\ttheta_{0}$ & \\
		\qquad{\it initialisation:} \\ 
		\qquad\quad\vline \quad Define the initial condition for the ansatz $\ket{\phi(\ttheta_{0})}$ & \\
		\qquad\quad\vline \quad Define the Hamiltonian $\widehat{\Hh}$ and decompose it as $\widehat{\Hh}=\sum_{i}\lambda_i h_i$ & \\
		\qquad{\it end} \\
		\qquad{\it while $\tau<\sigma^2T$:} \\	
		\qquad\quad\vline \quad Compute the matrix~$\Am(\tau)$ & (QC)\\
		\qquad\quad\vline \quad {\it if $\widehat{\Hh}$ is time-dependent:}\\	
		\qquad\quad\vline \quad \vline \ \  Update the Hamiltonian $\widehat{\Hh}(\tau)$ and its decomposition $\widehat{\Hh}(\tau)=\sum_{i}\lambda_i h_i$ & \\
		\qquad\quad\vline \quad {\it end} \\
		\qquad\quad\vline \quad Compute the vector~$\Cm(\tau)$ & (QC)\\				 				
		\qquad\quad\vline \quad Update the new values of $\ttheta$ & (CC)\\				 				
		\qquad{\it end}\\
		\qquad \textbf{output} $\rightarrow$ Price of the underlying derivative (in the original coordinates)&  \\		
		\hline
		\hline
	\end{tabular}
	\caption{Hybrid quantum-classical algorithm.}
	\label{Tab:Algo2}
\end{table}

The current methodology only checks the quantum approximation accuracy at the payoff level, i.e. 
verifying $\|\ket{\phi(\ttheta_{0})} - \ket{\psi(0))}\| = \Oo(\eps)$, for some tolerance~$\eps$,
assuming the error does not propagate much along the grid.
This assumption may fail when the price $\ket{\psi(\tau)}$ varies significantly.
One computational issue of the scheme is the resolution of the high-dimensional minimisation problem~\eqref{Eq:OptPr}. 
Other imaginary-time evolution algorithms~\cite{Motta2019} bypassing this optimisation might provide an efficient alternative for large-scale applications.
We leave these two issues for further investigations.

\section{Numerical Examples} \label{Sec:NE}
We now employ the imaginary time evolution technique above to compute the prices of two financial derivatives,
and we show that the ansatz circuit in Figure~\ref{Fig:AnsatzCircuit} is able to reconstruct the complete price evolutions accurately. 
This configuration leads to a quantum circuit, fully described in Appendix~\ref{Sec:AppAns}, 
composed of~$25$ ~$\Rm_y$ gates, resulting in $\ttheta_{\tau} \in \RR^{25}$ for each~$\tau$. 

\subsection{European Call Option} \label{Sub:Eur}
\begin{figure}[!h]
	\centering
	\includegraphics[scale=0.3]{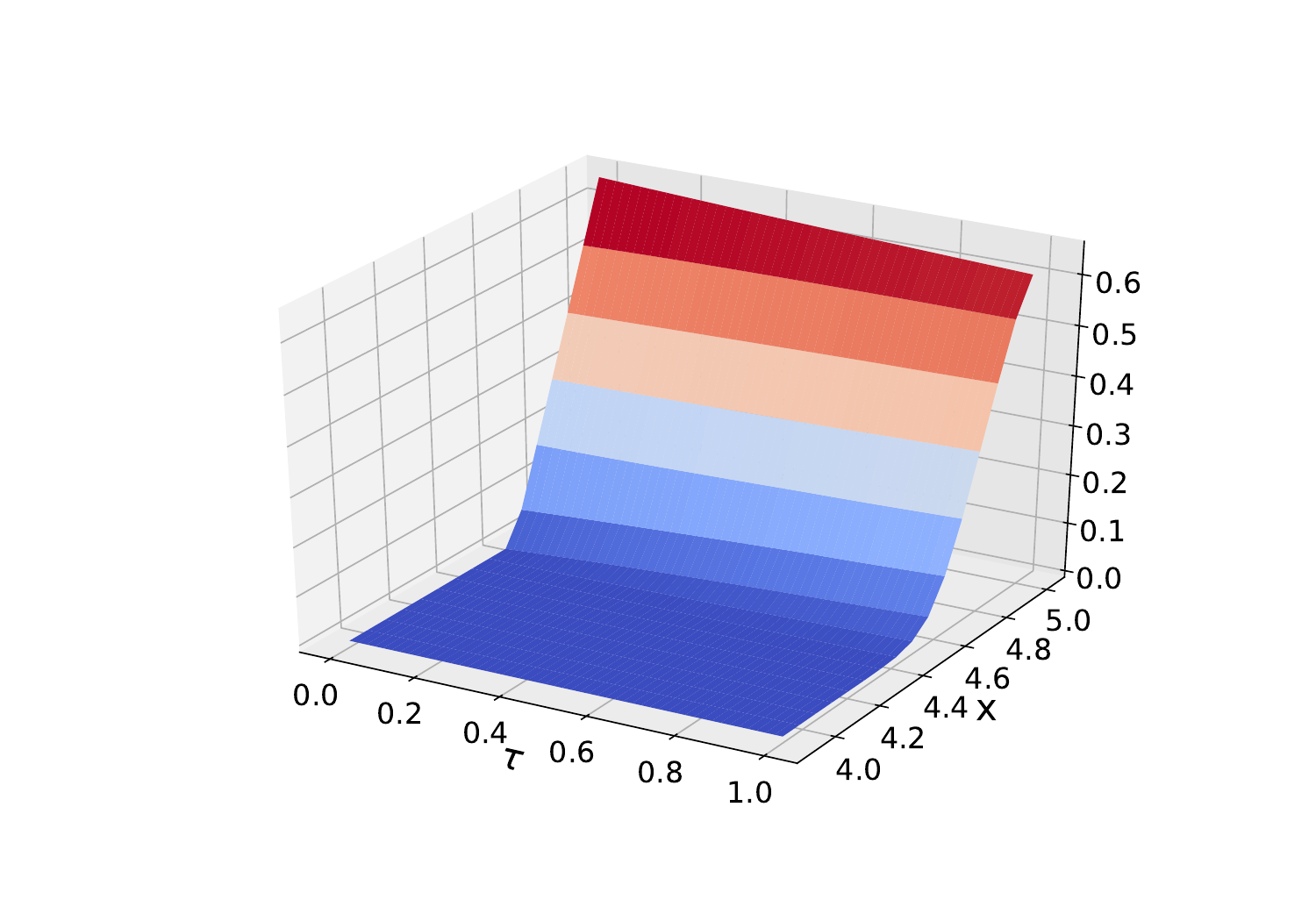}
	\includegraphics[scale=0.3]{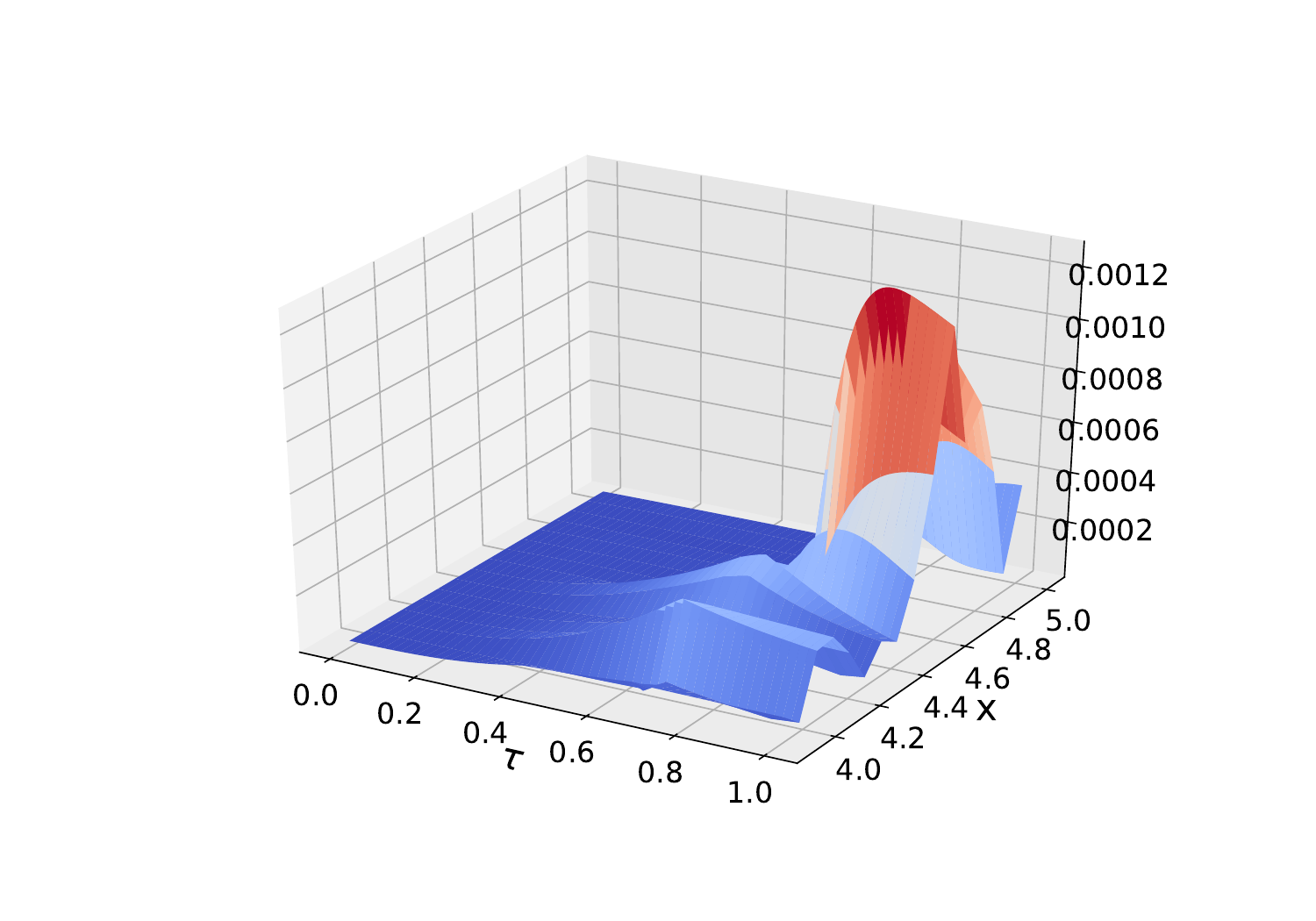}\\
	\includegraphics[scale=0.26]{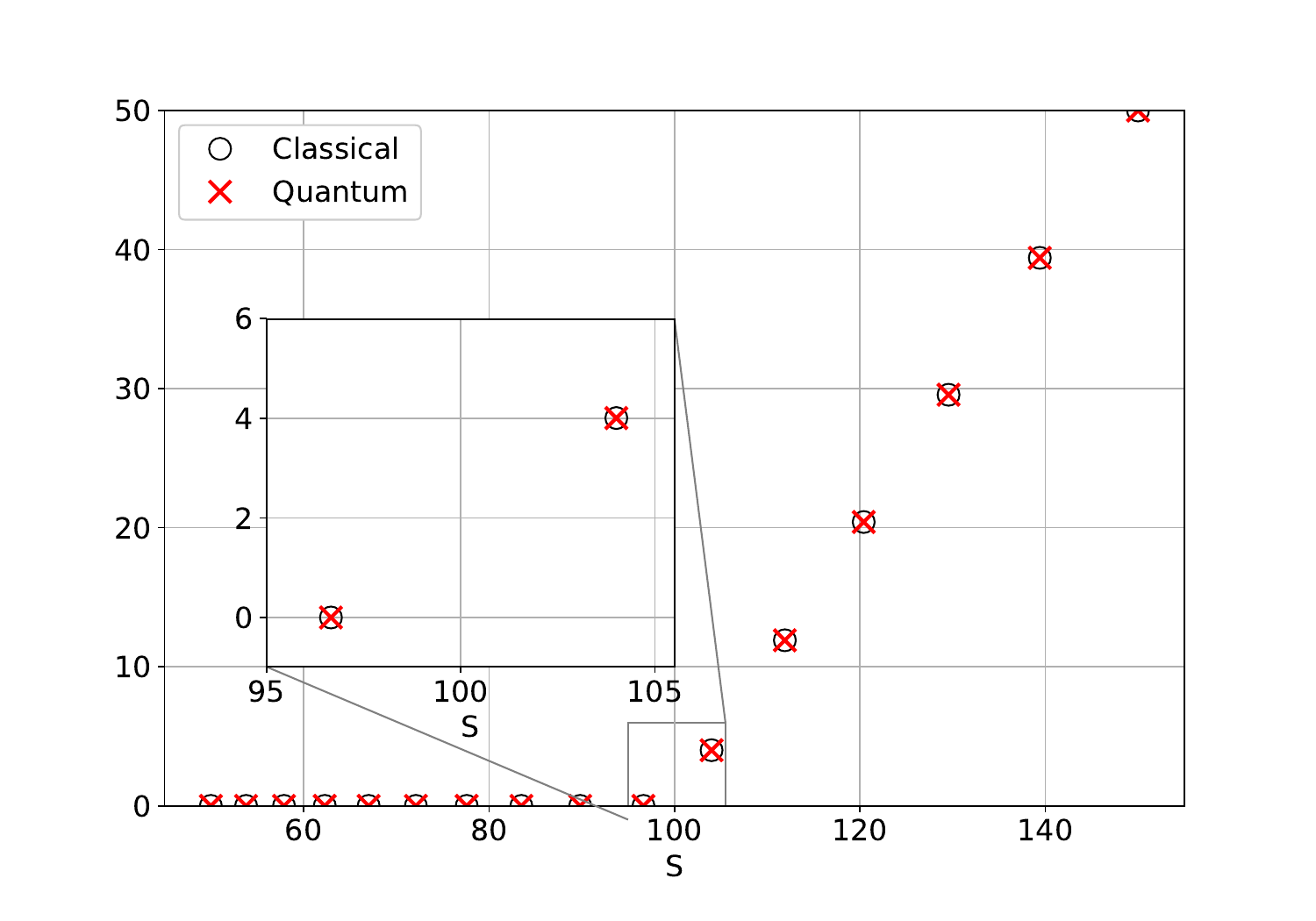}
	\includegraphics[scale=0.26]{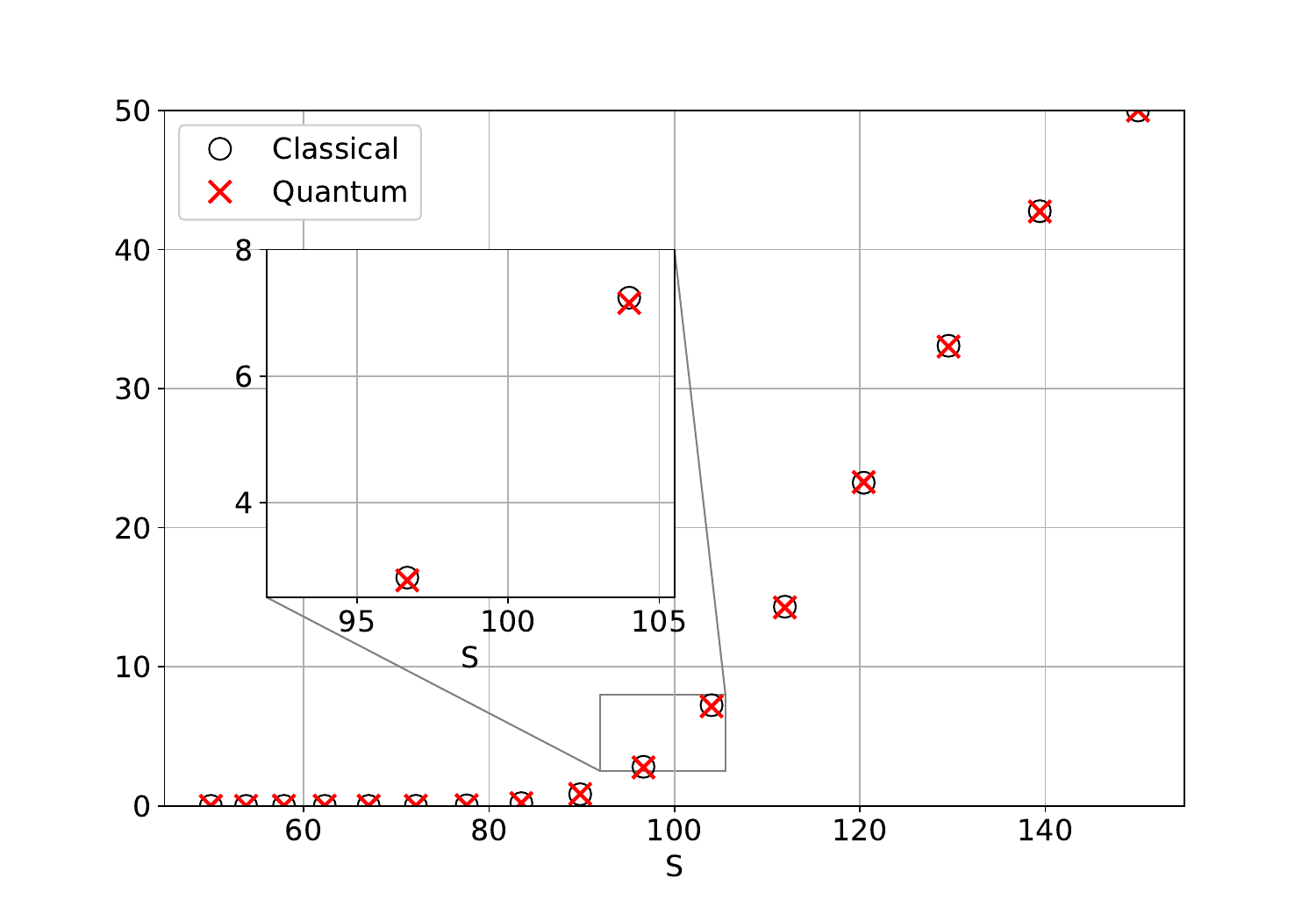}
	\caption{Top row: European prices using the quantum algorithm (left), and errors (right)
$\left\| \ket{\psi(\tau)} - \ket{\phi(\ttheta_{\tau})} \right\|$, with $n_T=500$ time steps.
Bottom row: Prices of European Calls using the hybrid quantum-classical algorithm at maturity (left) and at inception (right).
}
\label{Fig:European_solutions}
\end{figure}

In the Black-Scholes model~\eqref{Eq:BlackScholes} with $\sigma=20\%$, $S_0=K=100$, $T=1$,  and zero interest rate,
we discretise the state space on logarithmic scale on an equidistant grid $[x_{\min}, x_{\max}]\approx[3.9,5]$
(or $[S_{\min}, S_{\max}]=[50,150]$). 
With four qubits, the discretisation represents $\ket{\psi}$ using $2^4=16$ points, 
where the states $\ket{\out}=\ket{0000}$ and $\ket{\out}=\ket{1111}$
represent the solution respectively at~$x_{\min}$ and  ~$x_{\max}$. 
The evolution of $\ket{\phi(\ttheta_{\tau})}$ from expiry down to inception is computed using the approach described in Section~\ref{Sec:IT}. 
The Hamiltonian is $\widehat{\Hh}=\frac{1}{2}\partial_{xx}$ (Section~\ref{sec:Schrodinger}),  
and we discretise it using second-order finite differences
$$\frac{1}{2\Delta_{x}^2}
\begin{pmatrix}
		- 2b\Delta_{x}^2 & 0 & 0 & 0  & \dotsb & 0 & 0 & 0 \\
		1 & -2 & 1 & 0 & \cdots & 0 & 0 & 0\\  
		0 & 1 & -2 & 1 & \cdots & 0 & 0 & 0\\  
		\vdots & \vdots & \vdots & \vdots & \ddots & \vdots & \vdots & \vdots \\
		0 & 0 & 0 & 0 & \cdots & 1 & -2 & 1 \\
		0 & 0 & 0 & 0 & \cdots & 0 & 0 &-2b\Delta_{x}^2
\end{pmatrix},
$$
where $\Delta_{x}$ is the discretisation step in space, and the first and top rows of the matrix correspond 
to the behaviour at~$x_{\min}$ and~$x_{\max}$, where~$b$ comes from the change of variables in~\eqref{Eq:Heat}. 
Once~$\widehat{\Hh}$ is constructed, the evolution of~$\ttheta_{\tau}$ is obtained from the Euler scheme~\eqref{Eq:Eu_evo}. 
We split~$[0,T]$ into~$n_T$ steps, 
and compute~$\Am$ and~$\Cm$ in~\eqref{Eq:Eu_evo}. 
Since the underlying linear system is ill-conditioned, we use~\eqref{Eq:ThetaDot}  
to compute~$\dot{\ttheta}$ in~\eqref{Eq:LinSys} with a cutoff ratio of~$10^{-8}$. 
Figure~\ref{Fig:European_solutions} plots~$\ket{\phi(\ttheta_{\tau})}$ using the ansatz circuit in Figure~\ref{Fig:AnsatzCircuit}
and the difference between this and the price using a classical algorithm. 
The small errors $\left\| \ket{\psi(\tau)} - \ket{\phi(\ttheta_{\tau})} \right\|$ 
show that the proposed ansatz is able to reconstruct the evolution of~$\ket{\psi(\tau)}$ very accurately.
We also compare the expected solutions to those obtained from the simulation of the quantum algorithm at maturity and at inception, 
in the original coordinate system~\eqref{Eq:BS_1}. 
The scaling parameter~$\gamma(\tau)$ in~\eqref{Eq:Im_Time} is straightforward to compute since~$V(t,S_{\max})$ is known. 
Therefore, it is possible to determine $u(\tau, x_{\max})$ and, consequently, 
the scaling parameter between $\ket{\phi(\ttheta_{\tau})}$ and $u(\tau, x)$ in~\eqref{Eq:Im_Time} is known.  
We list the initial and final values of~$\ttheta$ in Appendix~\ref{Sec:AppITC}.

\newpage
\subsection{Asian Call Option}
\begin{figure}[!h]
	\centering
	\includegraphics[scale=0.3]{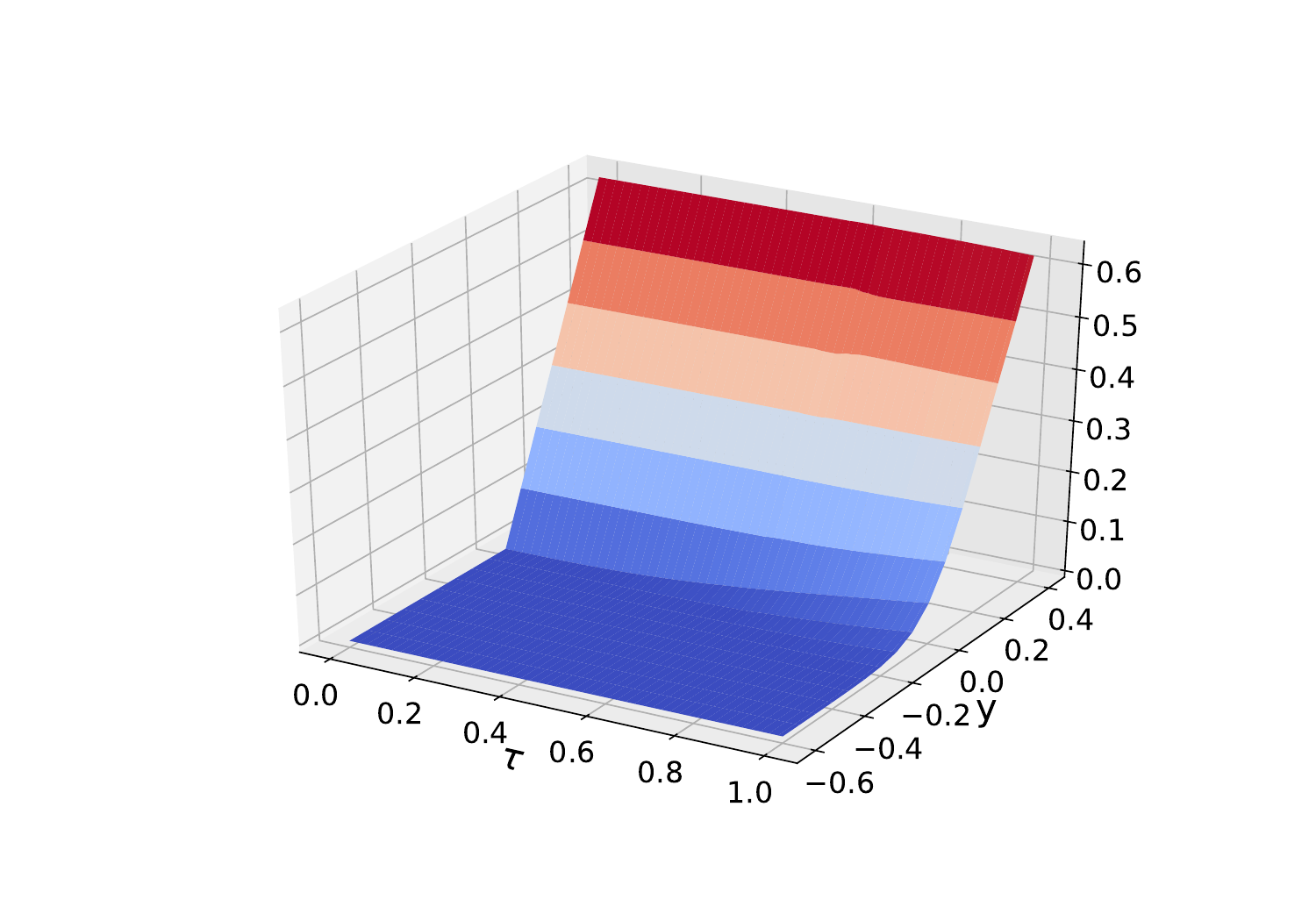}
	\includegraphics[scale=0.3]{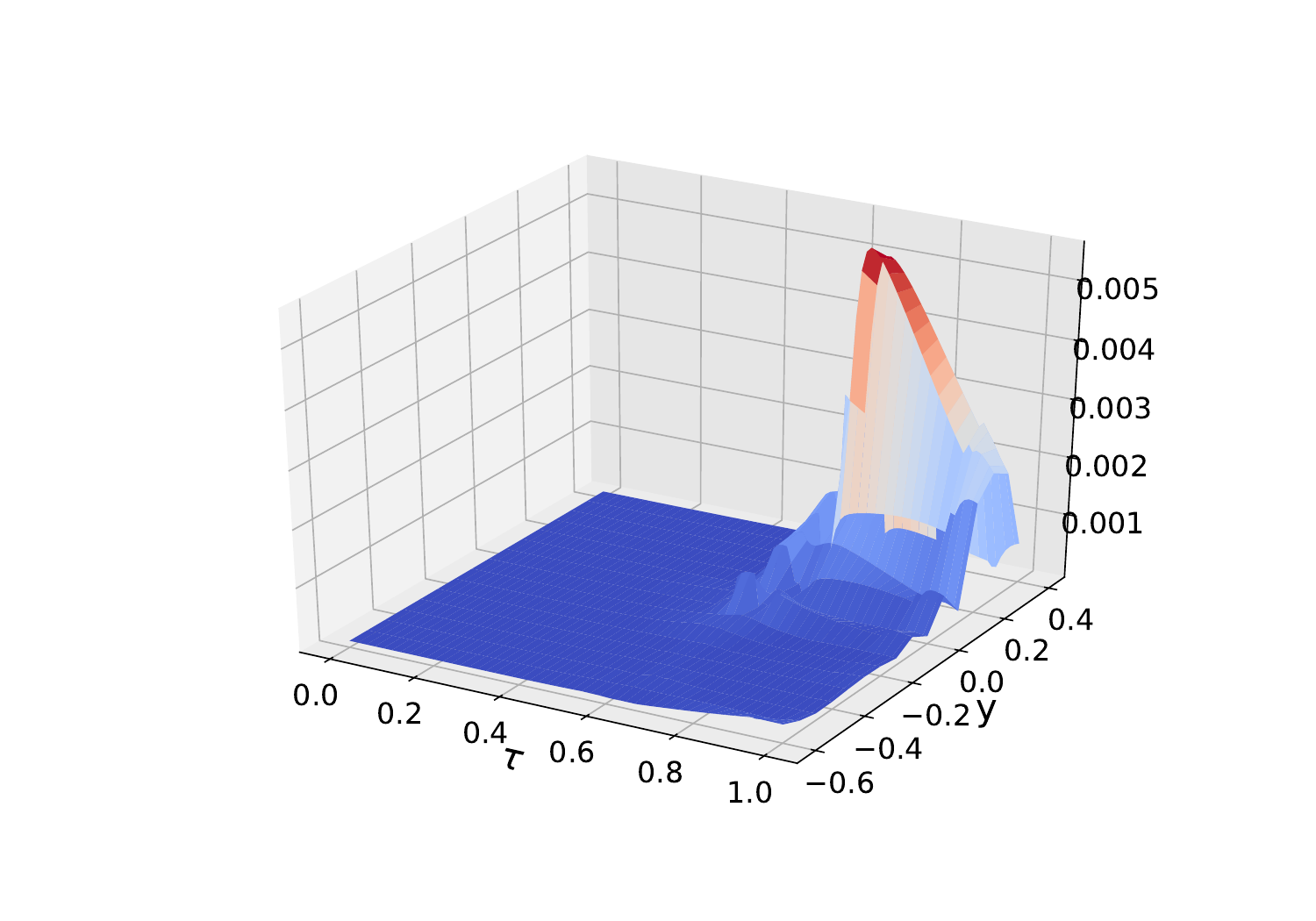}\\
	\includegraphics[scale=0.26]{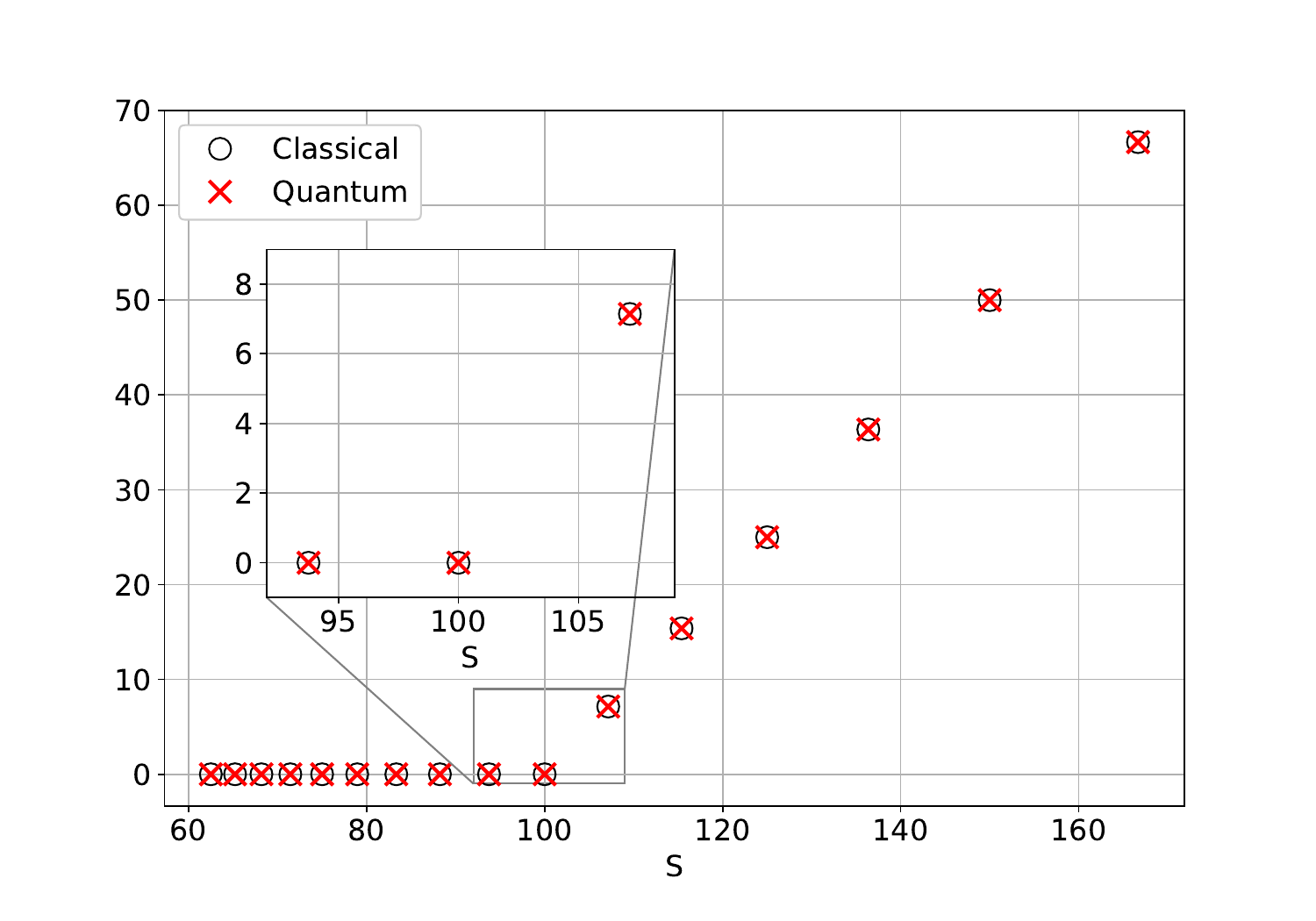}
	\includegraphics[scale=0.26]{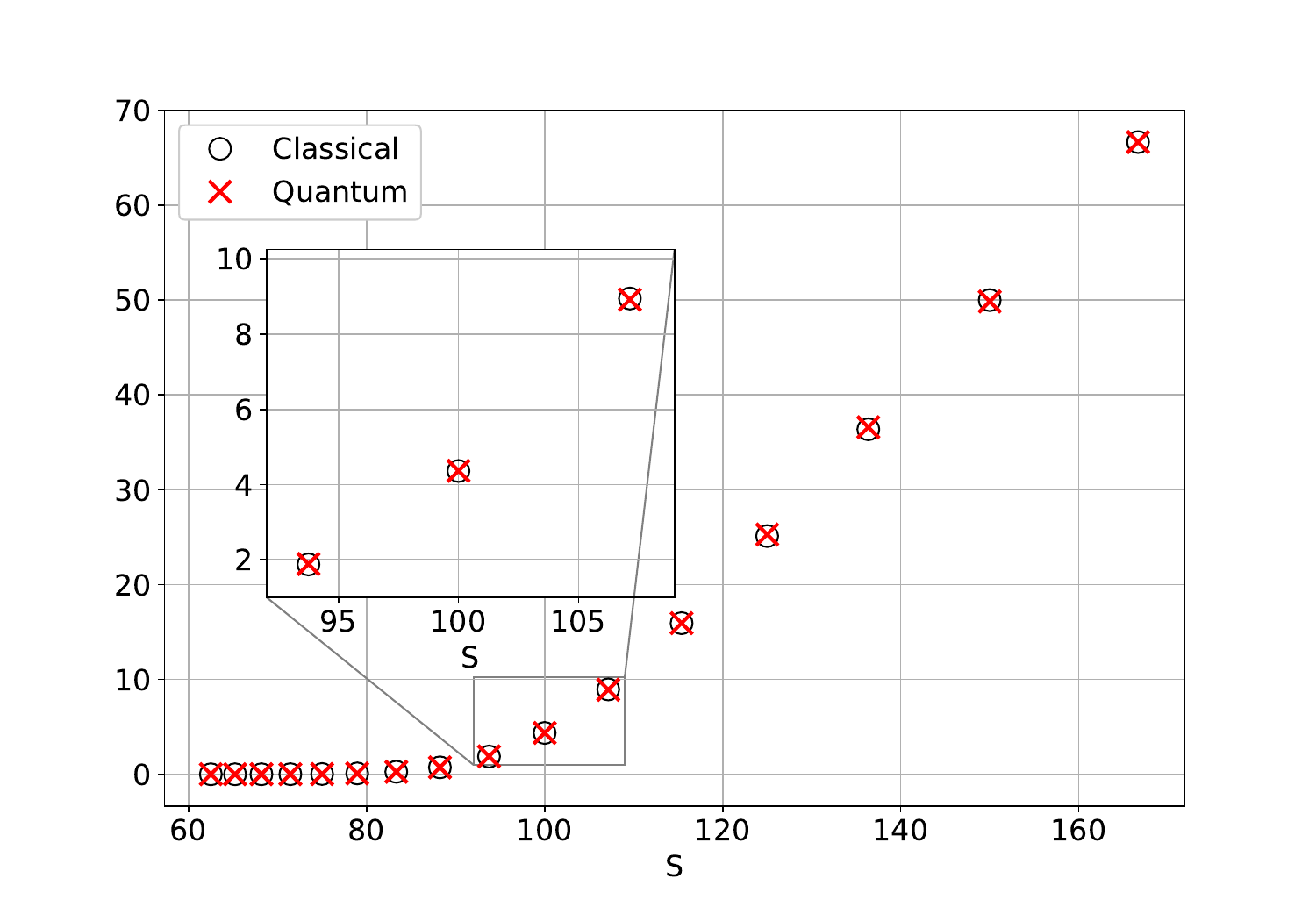}
	\caption{Same plots as in Figure~\ref{Fig:European_solutions} but in the Asian option case.}
	\label{Fig:Asian_solutions}
\end{figure}

We use the same parameters as in Section~\ref{Sub:Eur}, and the ansatz circuit in Figure~\ref{Fig:AnsatzCircuit}. 
The system~\eqref{Eq:Vecer} is computed on the equidistant grid $[y_{\min}, y_{\max}]=[-0.5, 0.4]$,
with the time-dependent Hamiltonian $\widehat{\Hh}(\tau)$ discretised as
$$
\frac{1}{2\Delta_{y}^2}
\begin{pmatrix}
		0 & 0 & 0  & \dotsb & 0 & 0 & 0 \\
		(q(\tau) - y_2)^2 & -2(q(\tau) - y_2)^2 & (q(\tau) - y_2)^2 & \dotsb & 0 & 0 & 0\\  
		\vdots & \vdots & \vdots & \ddots & \vdots & \vdots & \vdots \\
		0 & 0 & 0 & \dotsb & (q(\tau) - y_{15})^2 & -2(q(\tau) - y_{15})^2 & (q(\tau) - y_{15})^2 \\
		0 & 0 & 0 & \dotsb & 0 & 0 & 0 
\end{pmatrix},
$$
where $\{y_{\min}, y_2, \cdots, y_{15}, y_{\max}\}$ is the discretised space grid and~$\Delta_{y}$ the discretisation step. 
The first and last rows of the matrix are null since~$\ket{\psi(\tau)}$ is assumed constant at~$y_{\min}$ and~$y_{\max}$. 
Figure~\ref{Fig:Asian_solutions} displays the results of the classical and the quantum simulations in the original price coordinates. 
Again, since the solution $Q(\tau,y)$ in~\eqref{Eq:Vecer} is known at~$y_{\max}$, 
the corresponding scaling parameter~$\gamma(\tau)$ is trivial to compute. 
We also plot the payoffs as before;
again, the respective errors due to the quantum approximation of the initial conditions are negligible. 
The quantum approximation obtained from the ansatz accurately matches the expected values obtained by classical simulation. 
The initial and final values of~$\ttheta$ are available in Appendix~\ref{Sec:AppITC}.

\section{Conclusions and Outlook} \label{Sec:Conc}
Quantum computers are now part of the daily headlines in technology-related and financial news, 
and promise a new area where heavy computations can be carried out in the blink of an eye,
relegating slow pricing and onerous calibration to the Middle Ages.
While this promise is appealing, we are still far from it, and actual quantum computers are still in their infancy, 
still deeply affected by decoherence.
On the software side, current quantum algorithms require a large number of qubits together with precise gate implementations, 
which may limit their applications in the near future. 
We put forward here a hybrid quantum-classical algorithm, originating in computational chemistry, to price financial derivatives. It solves the Black-Scholes PDE by firstly transforming it into the Heat Equation and then converting the option price to a Quantum-State which is effectively a wave-function. The wave function is solved by the Hybrid Quantum and Classical Algorithm where a Quantum Circuit of imaginary time evolution is build with the help of McLachlan's invariance Principle. The strategy is based on the PDE representation of the pricing problem, and the link between the latter and its Schr\"{o}dinger counterpart from quantum mechanics. 
The results show that a shallow quantum circuit is able to represent European and Asian Call option prices accurately, 
and therefore suggest that this approach is a promising candidate for the application of quantum computing in Finance.

The main challenge of the present methodology is the requirement for an ansatz circuit and the corresponding solution of an optimisation problem. More work is needed in the future to design an efficient ansatz for more complex financial products, 
or in the development of an ansatz-free approach. 
One of the most promising application of this technique is for basket options, based on several stocks, and hence multidimensional. 
Classical PDE methods suffer from the so-called curse of dimensionality, making such pricing problem cumbersome, or at least computationally intensive.
Since its mathematical formulation is similar to the simulation of large-scale quantum mechanical systems governed by systems of Schr\"{o}dinger equations, 
we believe that a modified version of our algorithm is applicable there, leveraging the power of quantum computing,
and we leave this investigation for the next step.


\appendix

\section{Initial and terminal conditions of $\ttheta$} \label{Sec:AppITC}
\begin{table}[ht!]
	\centering
	\begin{tabular}{c | c cc c c }
		\hline
		\hline
		\multicolumn{1}{c}{} & \multicolumn{2}{c}{European Call Option} &  & \multicolumn{2}{c}{Asian Call Option} \\
		\hline
		\multicolumn{1}{c}{} & $\tau=0$ & $\tau= \sigma^2T$ & & $\tau = 0$ & $\tau=\sigma^2T$\\ 
		\hline
		$\theta^1$ & 3.142 & 3.203 & & 3.141 & 3.458\\
		$\theta^2$ & 4.173 & 4.149 & & 3.387 & 3.153\\
		$\theta^3$ & 1.392 & 2.278 & & 1.282 & 1.108\\
		$\theta^4$ & 3.713 & 3.512 & & 0.927 & 0.683\\
		$\theta^5$ & 2.399 & 2.400 & & 4.946 & 4.858\\
		$\theta^6$ & 0.935 & 0.512 & & 0.257 & 0.130\\  
		$\theta^7$ & 2.196 & 2.578 & & 2.937 & 4.062\\  
		$\theta^8$ & 5.014 & 5.024 & & 6.283 & 6.759\\ 
		$\theta^9$ & 2.736 & 2.405 & & 4.304 & 3.993\\ 
		$\theta^{10}$ & 1.477 & 2.406 & & 1.632 & 2.423\\ 
		$\theta^{11}$ & 4.472 & 4.271 & & 5.103 & 4.859\\
		$\theta^{12}$ & 3.415 & 3.375 & & 4.959 & 4.824\\ 
		$\theta^{13}$ & 6.283 & 6.331 & & 4.409 & 3.063\\ 
		$\theta^{14}$ & 4.244 & 4.224 & & 0.592 & 0.236\\  
		$\theta^{15}$ & 4.711 & 4.347 & & 6.283 & 6.599\\ 
		$\theta^{16}$ & 0.717 & 0.510 & & 0.133 & -0.693\\ 
		$\theta^{17}$ & 1.741 & 2.357 & & 3.957 & 3.923\\ 
		$\theta^{18}$ & 1.158 & 0.956 & & 0.334 & 0.090\\
		$\theta^{19}$ & 2.531 & 2.443 & & 0.489 & 2.035\\
		$\theta^{20}$ & 5.705 & 5.232 & & 4.940 & 4.582\\
		$\theta^{21}$ & 3.525 & 3.184 & & 0.909 & 0.935\\ 
		$\theta^{22}$ & 4.582 & 4.716 & & 3.141 & 3.118\\ 
		$\theta^{23}$ & 2.465 & 2.706 & & 5.993 & 4.807\\ 
		$\theta^{24}$ & 0.098 &-1.154 & & 2.807 & 3.324\\ 
		$\theta^{25}$ &	5.018 & 4.817 & & 2.783 & 2.539\\  
		\hline
	\end{tabular}
	\caption{Values of $\ttheta$ at initial and terminal times.}
	\label{Tab:theta_results}
\end{table}

\newpage

\section{Complete ansatz circuit} \label{Sec:AppAns}
\begin{figure}[!h]
	\centering
	\includegraphics[scale=2.3, angle=90]{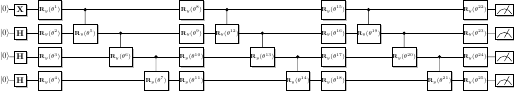}
	\caption{Ansatz circuit explained in Section~\ref{Sec:NE}. The circuit consists of~25~$\Rm_y$ gates.} 
	\label{Fig:Complete_ansatze}
\end{figure}


\end{document}